\documentclass[12pt]{article}
\usepackage{amssymb,amsmath}
\title{Gravitational scattering of a quantum particle and the privileged coordinate system }
\author{A.I.Nikishov}

\begin{document}

\maketitle

\begin{abstract}
 In gravitational scattering the  quantum
particle probes the Fourier-transforms of a metric. I evaluate  the Fourier-transforms of Schwarzschild metrics in standard, harmonic and other coordinate systems in linear and $G^2-$approximations. In general different coordinate systems lead to different scattering. This opens up the possibility to choose the privileged coordinate system  which  should lead to scattering in agreement with experiment.
\end{abstract}

In general relativity all permissible coordinate system are equivalent. Yet as V.Fock insisted there should be the privileged coordinate system and the theory should give the unique solution, see \S 93 in [1]. We note here that the theory of sources naturally gives unique solution.
As quantum scattering probes the Fourier transform of a metric, the idea is to use it in choosing the privileged coordinate system. For simplicity reason I assume that the test particle has 0-spin and interact only gravitationally by one graviton exchange. In case of other spins similar results are expected. Then the scattering particle  may be a hypothetical dark matter particle, graviton or neutrino.
  
The following notation is used
$$
g_{ik}=\eta_{ik}+h_{ik}^{(1)}+h_{ik}^{(2)}+\cdots,\quad h_{ik}^{(n)}\propto G^n,\quad 
h_{,l}=\frac{\partial h}{\partial x_l},\quad h=h^k{}_k, \quad \eta_{ik}={\rm diag}(-1,1,1,1).
$$
The Latin indexes run from 0 to 3, the Greek ones from 1 to 3. 
In each coordinate system we denote coordinates by the same $x^i$. This is because we use it only as an integration variables in obtaining Fourier-transforms.

We start with the simplest case of linear approximation to Schwarzschild metric and assume at first that the point-like particle is the source of gravitational field. The standard coordinate system in
linear and $G^2$-approximation has the form
$$
 h^{(1)st}_{\alpha\beta}=\frac{2GMx_\alpha x_{\beta}}{r^3},\quad
 h^{(2)st}_{\alpha\beta}=\frac{(2GM)^2x_\alpha x_{\beta}}{r^4}.
                                                                                                     \eqno(1)
$$

see S.Weinberg [2], Ch 8, \S 2 and get  the necessary approximation of unnumbered expression after eq.(8.2.15)
In linear approximation  the harmonic and isotropic systems coinside
$h^{(1)har}_{ik}= h^{(1)iso}_{ik}=\frac{2MG}{r}\delta_{ik}$  and $ h^{(1)}_{00}=\frac{2MG}{r}$ is the same in 
all coordinate systems. Now 
$$
h_{\alpha\beta}^{(1)st}-h_{\alpha\beta}^{(1)iso}=-MG(\Lambda_{\alpha,\beta}^{(1)}+\Lambda_{\beta,\alpha}
^{(1)}).\quad \Lambda_{\alpha,\beta}^{(1)}=(\frac{\delta_{\alpha\beta}}{r}-\frac{x_{\alpha} x_{\beta}}{r^3}), \quad \Lambda_{\alpha}^{(1)}=\frac{x_{\alpha}}{r}.                                                                                                                                                                    \eqno(2)
$$
So the difference in the l.h.s. is a gauge function.

Now the scattering amplitude of two scalar particles exchanging by one graviton is proportional to
$$
T^{ij}(p,p')\frac{P_{ijkl}}{q^2}T^{kl}(k,k'),\quad T^{ij}(p,p')=\frac12(p^ip'^j+p^jp'^i)-\frac12\eta^{ij}(pp'+M^2),
$$
$$
P_{ijkl}=\frac12(\eta_{ik}\eta_{jl}+\eta_{il}\eta_{jk}-\eta_{ij}\eta_{kl}),\quad q=k'-k=p-p', \eqno(3)
$$ 
see for example  \S 4.3 in R.Feynman and others [3] (notice that in [3] $\eta_{ik}={\rm diag}(1,-1,-1,-1)$). It is easy to check that
$q_iT^{ij}(p,p')=0$  and similarly for $T^{kl}(k,k')$. This is due to the conservation law for energy-momentum
tensor.and it holds for the particle of any spin.

Next we assume that the particle with momentum $p, p^2=-M^2$ is a heavy one and the particle with momentum $k, k^2=-m^2$ is the probing one. Then
$$
T^{ij}(p,p')\frac{P_{ijkl}}{q^2}=\frac12\frac{M^2}{q^2}\delta_{kl}.
$$
Up to a factor which is not interested us now this is Fourier-transform of $h_{kl}^{(1)iso}$: 
$\int d^3xe^{i\vec q\cdot\vec x}h^{(1)iso}_{kl}(x)=4\pi\delta_{kl}2MG/q^2$. So the heavy particle becomes the source of the
Schwarzschild field in Hilbert (or Lorentz) gauge: $\bar h^{kl}{}_{,l}=(h^{kl}-\frac12\eta^{kl}h)_{,l}=0$.
The coordinate system in this gauge I consider as the privileged one [4]. In this linear approximation it
coinsides with harmonic system and agrees with Fock definition of privileged system.

Now we evaluate the Fourier-transforms of metrics in different coordinate system. We use the expression
$$
\int d^3xe^{i\vec q\cdot\vec x}\frac{x_{\alpha}x_{\beta}}{r^n}=\frac{q_{\alpha} q_{\beta}}{q^2}A(q)+\delta_{\alpha\beta}B(q).                                                  \eqno(4)
$$ 
Cases $n=3$ and $n=4$  admit the limit $b=0$ where $b$ is the radius of the ball of matter. So in these cases we put $b=0$. The functions $A(q)$ 
and $B(q)$ can be obtained as fallows. We put $\alpha=\beta $. Then
$$
\int d^3xe^{i\vec q\cdot\vec x}\frac{1}{r^{n-2}}= A(q)+3B(q).                                                                                                                           \eqno(5)
$$
Next we assume temporarily that $q_{\alpha}=(0,0,q)$ and use $x_3=rt, t=\cos \theta$. Then from (4) we have
$$
\int d^3xe^{i\vec q\cdot\vec x}\frac{t^2}{r^{n-2}}= A(q)+B(q).                                                                                                                         \eqno(6)
$$
To evaluate the l.h.s. we twice differentiate  over $a$ the relation $\int _{-1}^{1}dte^{iat}=\frac2a\sin a$. Thus we get
$$
\int _{-1}^{1}dte^{iat}t^2=2[(\frac1a-\frac{2}{a^3})\sin a+\frac{2}{a^2}\cos a]. \eqno(7)
$$ 
Using this equation we find for the l.h.s. of (6) for $n=3$ and $n=4$
$$
\int d^3xe^{i\vec q\cdot\vec x}\frac{t^2}{r}=-\frac{4\pi}{q^2}, \quad
\int d^3xe^{i\vec q\cdot\vec x}\frac{t^2}{r^{2}}=0.                              \eqno(8)
$$

For the l.h.s. of (5) putting $ \alpha=\beta$ and $n=3$ we obtain just as in Coulomb scattering
$$
\int d^3xe^{i\vec q\cdot\vec x}\frac{1}{r}=\frac{4\pi}{q^2}= A(q)+3B(q).                        \eqno(9)
$$
On the other hand from (6) and the first equation in (8) for $n=3$ we have
$$
-\frac{4\pi}{q^2}= A(q)+B(q).                                                 \eqno(10)
$$
Now from (9) and (10) it follows that $ A(q)=-8\pi/q^2, B(q)=4\pi/q^2$. Hence equation (4) for $n=3$ gives
$$
\int d^3xe^{i\vec q\cdot\vec x}\frac{x_{\alpha}x_{\beta}}{r^3}=\frac{4\pi}{q^2}(-2\frac{q_{\alpha} q_{\beta}}{q^2}+\delta_{\alpha\beta}).                                                  \eqno(11)
$$

In a similar manner we obtain
$$
\int d^3xe^{i\vec q\cdot\vec x}\frac{x_{\alpha}x_{\beta}}{r^4}=\frac{\pi^2}{q}(-\frac{q_{\alpha} q_{\beta}}{q^2}+\delta_{\alpha\beta}).                                                  \eqno(12)
$$
Putting here $\alpha=\beta$ we find
$$
\int d^3xe^{i\vec q\cdot\vec x}\frac{1}{r^2}=2\frac{\pi^2}{q}.                                                                                                                                          \eqno(13)
$$

Now from (9) and (11) for $\Lambda^{(1)}_{\alpha,\beta}$ (defined in (2)) we get
$$
\int d^3xe^{i\vec q\cdot\vec x}\Lambda_{\alpha,\beta}^{(1)}=8\pi\frac{q_{\alpha} q_{\beta}}{q^4}, \eqno(14)
$$
similarly from (12) and (13) we have
$$
\int d^3xe^{i\vec q\cdot\vec x}\Lambda_{\alpha,\beta}^{(2)}=2\pi^2\frac{q_{\alpha }q_{\beta}}{q^3},\quad
\Lambda^{(2)}_{\alpha,\beta}= (\frac{x_{\alpha}}{r^2})_{,\beta}=
\frac{\delta_{\alpha\beta}}{r^2}-\frac{2x_{\alpha} x_{\beta}}{r^4}.     \eqno(15)                             
$$
The equations (2) and (14) mean that $h_{ik}^{(1)st}$ and $h_{ik}^{(1)har}$ (which is equal $h_{ik}^{(1)iso}$) lead to the same scattering due to the conservation law of energy-momentum tensor of the probing particle: $q_{\alpha}T_{\alpha\beta}(k,k')=0$. 
We remind here that $q_0=0$ when $M>>k_0$. 

Now in $G^2$-approximation we have$$
h_{\alpha\beta}^{(2)har}=G^2M^2\left(\frac{x_{\alpha}x_{\beta}}{r^4}+\frac{\delta_{\alpha\beta}}{r^2}\right), \quad
 h_{\alpha\beta}^{(2)iso}=\frac{3M^2G^2}{2r^2},                                      \eqno(15a),          
$$
$$
h_{00}^{(2)har}=h_{00}^{(2)iso}=-\frac{2M^2G^2}{r^2}, \quad h_{\alpha\beta}^{(2)har}-h_{\alpha\beta}^{(2)iso}
=-\frac12M^2G^2\Lambda_{\alpha,\beta}^{(2)}.                                           \eqno(15b)
$$
The first equation in (15a) can be obtained from equation (8.2.15) in [2], the second from Problem 4 in \S 100 in [8].

Now from (15) and the last equation in (15b) it follows that $h_{\alpha\beta}^{(2)iso}$ and 
$h_{\alpha\beta}^{(2)har}$ also lead to the same scattering. In other words  the 
differences in metrics by a gauge function does not affect the scattering. 

On the other hand the difference
$$
h_{\alpha\beta}^{(2)st}-h_{\alpha\beta}^{(2)har}=G^2M^2(\frac{3x_{\alpha}x_{\beta}}{r^4}-\frac{\delta_{\alpha\beta}}{r^2})                                                                             \eqno(16)
$$
it is not a gauge function and the scattering is different in these coordinate systems in  $G^2$- terms.
Indeed, in case of harmonic coordinate  the contribution to scattering amplitude is proportional to
$$
\int d^3x\exp{i\vec q\cdot \vec x}[T_{00}(k,k')h^{(2)har}_{00}+T_{\alpha\beta}(k,k')h^{(2)har}_{\alpha\beta}]=G^2M^2\frac{\pi^2}{q}
[\frac52k_0^2-\frac72 \vec k\cdot\vec k'-\frac{13}{2}m^2].                         \eqno(17)]
$$
The same result we get if instead of $h_{\alpha\beta}^{(2)har}$ we use $h_{\alpha\beta}^{(2)iso}$.
In case of standard system instead of (17) we have
$$
\int d^3x\exp{i\vec q\cdot \vec x}[T_{\alpha\beta}(k,k')h^{(2)st}_{\alpha\beta}=G^2M^2\frac{\pi^2}{q}
[6k_0^2-2 \vec k\cdot\vec k'-6m^2],\quad h_{00}^{(2)st}=0.                                                          \eqno(18)]
$$
We see that (17) and (18) are different.

Now we take into account the finiteness of radius $b$ of the ball of matter.
Then for the privileged coordinate system
$$
h_{\alpha\beta}^{(1)priv}=\left\{\begin{array}{cc}
\frac{MG}{b}(3-\frac{r^2}{b^2})\delta_{\alpha\beta}, \quad  r<b,\\
\frac{2MG}{r}\delta_{\alpha\beta},                   \quad  r>b,
                                 \end{array}\right.                                     \eqno(19)
$$
and for the standard system
$$
h_{\alpha\beta}^{(1)st}=\left\{\begin{array}{cc}
2MG\frac{x_{\alpha}x_{\beta}}{b^3}, \quad  r<b, \\
2MG\frac{x_{\alpha}x_{\beta}}{r^3},   \quad  r>b.
                                 \end{array}\right.                                     \eqno(20)
$$

Now for the Fourier-transforms using equations (41)and (42) below we obtain
$$
\int_{r<b}d^3xe^{i\vec q\cdot\vec x}h_{\alpha\beta}^{(1)priv}=2MG\frac{4\pi}{q^2}\{\frac{3}{a^2}(\frac{\sin a} {a}-\cos a)-\cos a\}\delta_{\alpha\beta},                                   \eqno(21)
$$
$$
\int_{r>b}d^3xe^{i\vec q\cdot\vec x}h_{\alpha\beta}^{(1)priv}=2MG\frac{4\pi}{q^2}\delta_{\alpha\beta}
\cos a
                                                                                         \eqno(22)
$$
The sum of (21) and (22) is
$$
\int_{0<r}d^3xe^{i\vec q\cdot\vec x}h_{\alpha\beta}^{(1)priv}=2MG\frac{4\pi}{q^2}\frac{3}{a^2}(\frac{\sin a}
{a}
-\cos a)\delta_{\alpha\beta},                                   \eqno(23)
$$

Similarly for the standard system
$$
\int_{r<b}d^3xe^{i\vec q\cdot\vec x}h_{\alpha\beta}^{(1)st}=2MG\frac{4\pi}{q^2}
\{\frac{q_{\alpha}q_{\beta}}{q^2}[(\frac{15}{a^2}-1)\cos a+
$$
$$
(\frac{6}{a}-\frac{15}{a^3})\sin a]  +\delta_{\alpha\beta}[-\frac{3}{a^3}\cos a+(-\frac1a+\frac{3}{a^3})\sin a
]\},                                                                   \eqno(24)
$$
$$
\int_{r>b}d^3xe^{i\vec q\cdot\vec x}h_{\alpha\beta}^{(1)st}=2MG\frac{4\pi}{q^2}\{\frac{q_{\alpha}q_{\beta}}{q^2}(\cos a-\frac{3\sin a}{a})  +\delta_{\alpha\beta}\frac{\sin a}{a}\}.
                                                                                         \eqno(25)
$$
The sum of (24) and (25) is
$$
\int_{0<r}d^3xe^{i\vec q\cdot\vec x}h_{\alpha\beta}^{(1)st}=2MG\frac{4\pi}{q^2}\{  \frac{q_{\alpha}q_{\beta}}{q^2}[\frac{15}{a^2}\cos a +(\frac3a-\frac{15}{a^3})\sin a]+ \delta_{\alpha\beta}[
 -\frac{3}{a^2}\cos a +\frac{3}{a^3}\sin a]
\}.                                                                                 \eqno(26)
$$
As terms containing $q_{\alpha}q_{\beta}$ do not contribute to scattering, it follows from (23) and (26)
that in our approximation  the privileged and standard coordinate systems lead to the same scattering:
$$
d\sigma=M^2G^2\left(\frac{k_0^2-m^2/2}{\vec k^2}\right)^2\frac{d\Omega}{\sin^4\frac{\theta}{2}}F^2(a),\quad
F(a)=\frac{3}{a^2}(\frac{\sin a}{a}-\cos a),\quad \left.F(a)\right|_{a<<1}=1-\frac{a^2}{10}+\cdots.     
                                                                                    \eqno(27)
$$
So for $a<<1$ we return to the scattering on a point-like source [5]. For finite $a$
there are oscillations in the angular distribution in (27) because $a=bq=b2|\vec k|\sin \theta/2$.

Finally we consider the case of finite radius of the matter ball in $G^2$-approximation. 
Now the theory of sources suggests that the privileged coordinate system is that one in which the gauge degrees of freedom are put to zero [4]. Assuming that the 3-graviton vertex is given by general relativity, such a system in $G^2$-approximation 
was obtained in [6]. It was also shown there that  the theory of sources leads to the appearance in external metric in  $G^2$-approximation of the term
$$
\varkappa M^2G^2b(\frac{x_{\alpha}x_{\beta}}{r^5}-\frac{\delta_{\alpha\beta}}{3r^3}).             \eqno(28)
$$ 
Here $b$ is the radius of the ball of matter, $\varkappa$ depends on the assumed form of the 3-graviton vertex and (contrary to general relativity, where the exterior metric does not depends on interior rigion) on the energy-momentum tensor of the matter.  It is of order unity. Despite the fact that (28) have the form of a gauge function it is generated by a source. Now we get the contribution to its  Fourier-transform from the exterior region. In equation (5) and (6) we put $n=5$. Using (7) we obtain
$$
\int_{b<r} d^3xe^{i\vec q\cdot\vec x}\frac{t^2}{r^3}=4\pi\int_a^{\infty} dx[-\frac{2\sin x}{x^4}+\frac{2\cos x}{x^3}+\frac{\sin x}{x^2}]= 4\pi[\frac23\frac{\sin x}{x^3}-\frac23\frac{\cos x}{x^2}-\frac13\frac{\sin x}{x}
$$
$$
+\left.\frac13\int\frac{\cos x}{x}dx]\right|_a^{\infty}=4\pi[-\frac23\frac{\sin a}{a^3}+\frac23\frac{\cos a}{a^2}+\frac13\frac{\sin a}{a}+\frac13\int_{a}^{\infty}\frac{\cos x}{x}dx].                                                                                                                                \eqno(29)
$$
Just  as in equation (8) and (9) we find for $B(q)$ and $A(q)$ in equations (4) and (5) for $n=5$
$$
B(q)=4\pi[\frac13(\frac{1}{a}+\frac{1}{a^3})\sin a-\frac13\frac{\cos a}{a^2}+\frac13\int_{a}^{\infty}\frac{\cos x}{x}dx],                                                                                                                                                                                                                                    \eqno(30)
$$
$$
A(q)=4\pi[-\frac{\sin a}{a^3}+\frac{\cos a}{a^2}], \quad a=qb,                                                                                                                              \eqno(31)
$$
and from here
$$
\int_{b<r} d^3xe^{i\vec q\cdot\vec x}(\frac{x_{\alpha}x_{\beta}}{r^5}-\frac{\delta_{\alpha\beta}}{3r^3})=
A(q)(\frac{q_{\alpha}q_{\beta}}{q^2}-\frac{\delta_{\alpha\beta}}{3})                                                                                                                              .\eqno(32)
$$

It will be seen below that this term can contribute to scattering only in more elaborate approximation, which probably include absorption in the matter ball or more then one graviton  exchange with gravitational source. 

As shown below in the considered approximation $h_{\alpha\beta}^{(2)priv}$ and $h_{\alpha\beta}^{(2)har}$
lead to the same scattering.
 According to the theory of sources (assuming that the nonlinear sources are given by general relativity) we have, see equations (18), (25a) and (19) in  [6]
$$
h_{\alpha\beta}^{(2)priv}=
(\frac{MG}{b})^2[\left(\frac{25}{4}-\frac{39}{19}\frac{r^2}{b^2}+\frac{23}{28}\frac{r^4}{b^4}\right)\delta_{\alpha\beta}-\frac{9}{5}\frac{x_{\alpha}x_{\beta}}{b^2}+
\frac{2}{7}\frac{r^2x_{\alpha}x_{\beta}}{b^4}], \quad  r<b\\   \eqno(33a)
$$
$$
h_{\alpha\beta}^{(2)priv}=(MG)^2[\frac{5\delta_{\alpha\beta}}{r^2}-\frac{7x_{\alpha}x_{\beta}}{r^4}+\frac{192}{35}b\left(\frac{x_{\alpha}x_{\beta}}{r^5}-\frac{\delta_{\alpha\beta}}{3r^3}\right)],                   \quad  r>b,
                                                                     \eqno(33b)
$$
$$
h_{00}^{(2)priv}=h_{00}^{(2)har}=h_{00}^{(2)iso}=
(\frac{MG}{b})^2[-\frac{15}{4}+\frac{3}{2}\frac{r^2}{b^2}+\frac14\frac{r^4}{b^4}], \quad  r<b\\ \eqno(34a)
$$
$$
h_{00}^{(2)priv}=h_{00}^{(2)har}=h_{00}^{(2)iso}=-2\frac{(MG)^2}{r^2},                   \quad  r>b,
                                 .                                     \eqno(34b)
$$
For harmonic $h_{\alpha\beta}^{(2)har}$ and isotropic $h_{\alpha\beta}^{(2)iso}$ we have, see see equations    (6) and (13) in [7]
$$
h_{\alpha\beta}^{(2)har}=\frac{M^2G^2}{b^2}\{\delta_{\alpha\beta}(\frac95-\frac{3r^2}{2b^2}-\frac{r^4}{4b^4})
+\frac{3x_{\alpha}x_{\beta}}{b^2}-\frac{2r^2x_{\alpha}x_\beta}{b^4}   \},\quad        r<b,                                                                                                                                                                                                       \eqno(35)
$$
$$
h_{\alpha\beta}^{(2)har}=M^2G^2(\frac{x_{\alpha}x_{\beta}}{r^4}+\frac{\delta_{\alpha\beta}}{r^2}),\quad r>b,                                                                                     \eqno(36)
$$
$$
h_{\alpha\beta}^{(2)iso}=\frac{M^2G^2}{b^2}(\frac{15}{4}-\frac{3r^2}{b^2}+\frac{3r^4}{b^4}), \quad r<b,\eqno(37)
$$
$$
h_{\alpha\beta}^{(2)iso}=\frac32\frac{M^2G^2}{r^2}\delta_{\alpha\beta}, \quad r>b.  \eqno(38)
$$
From (35) and (33a) we have
$$
h_{\alpha\beta}^{(2)har}-h_{\alpha\beta}^{(2)priv}=\frac{M^2G^2}{b^2}\{-4\delta_{\alpha\beta}+\frac{12}{5}(
\delta_{\alpha\beta}\frac{r^2}{b^2}+2\frac{x_{\alpha}x_{\beta}}{b^2})
-\frac47(\delta_{\alpha\beta}\frac{r^4}{b^4}-4\frac{r^2x_{\alpha}x_{\beta}}{b^4}\}, \quad r<b                                                                                                                \eqno(39)
$$
In the r.h.s. of (39) stand gauge functions.
Similarly from (36) and (33b)  we have
$$
h_{\alpha\beta}^{(2)har}-h_{\alpha\beta}^{(2)priv}=M^2G^2\{-4\frac{\delta_{\alpha\beta}}{r^2}+
8\frac{x_{\alpha}x_{\beta}}{r^4}-\frac{192}{35}b(\frac{x_{\alpha}x_{\beta}}{r^5}-
\frac13\frac{\delta_{\alpha\beta}}{r^3})   \}, r>b.                                              \eqno(40)      $$
As expected the right hand sides of (39) and (40) are the gauge function. This is because the  nonlinear sources  are the same.

Now we give building blocks for calculating the Fourier-transforms. For contributions from $r<b$:
$$
\int_{r<b}d^3xe^{i\vec q\cdot\vec x}\frac{1}{b^2}=\frac{4\pi}{q}(\frac{\sin a}{a^2}-\frac{\cos a}{a}),\quad a=qb, \eqno(41)
 $$
$$
\int_{r<b}d^3xe^{i\vec q\cdot\vec x}\frac{r^2}{b^4}=\frac{4\pi}{q}\{(\frac{3}{a^2}-\frac{6}{a^4})\sin a+(-\frac{1}{a}+\frac{6}{a^3})\cos a  \},                                                 \eqno(42),
$$

 $$
\int_{r<b}d^3xe^{i\vec q\cdot\vec x}\frac{r^4}{b^6}=\frac{4\pi}{q}\{(\frac{5}{a^2}-\frac{3\cdot4\cdot5}{a^4}+\frac{5!}{a^6})\sin a+(-\frac{1}{a}+\frac{4\cdot5}{a^3}-\frac{5!}{a^5})\cos a  \},                               \eqno(43),
$$
$$
\int_{r<b}d^3xe^{i\vec q\cdot\vec x}\frac{x_{\alpha}x_{\beta}}{b^4}=\frac{4\pi}{q}\{\frac{q_{\alpha}q_{\beta}}{q^2}[(\frac{6}{a^2}-
\frac{15}{a^4})\sin a+(-\frac{1}{a}+\frac{15}{a^3})\cos a]+
$$
$$
\delta_{\alpha\beta}[(-\frac{1}{a^2}+\frac{3}{a^4})\sin a-\frac{3}{a^3}\cos a]  \},                                                                                                                \eqno(44),
$$
$$
\int_{r<b}d^3xe^{\vec q\cdot\vec x}\frac{r^2x_{\alpha}x_{\beta}}{b^6}= \frac{4\pi}{q}\{\frac{q_{\alpha}q_{\beta}}{q^2}[(\frac{8}{a^2}-
\frac{3\cdot5\cdot7}{a^4}+\frac{2\cdot3\cdot5\cdot7}{a^6})\sin a+
$$
$$
(-\frac{1}{a}+
\frac{5\cdot7}{a^3}-\frac{2\cdot3\cdot5\cdot7}{a^5})\cos a]+\delta_{\alpha\beta}[(-\frac{1}{a^2}+\frac{3\cdot5}{a^4}-
$$
$$
\frac{2\cdot3\cdot5}{a^6})\sin a+(-\frac{5}{a^3}+\frac{2\cdot3\cdot5}{a^5})\cos a]   \},                                                                                                                 \eqno(45)
$$

For contributions from $r>b$ we have:
$$
\int_{b<r}d^3xe^{i\vec q\cdot\vec x}=\frac{4\pi}{q^2}\cos a,\quad 
\int_{b<r}d^3xe^{i\vec q\cdot\vec x}\frac{x_{\alpha}x_{\beta}}{r^3}=\frac{4\pi}{q^2}\{\frac{q_{\alpha}q_{\beta}}{q^2}(-\frac{3}{a}\sin a+\cos a)+
\delta_{\alpha\beta}\frac{\sin a}{a} \},                                                          \eqno(45a)                           
$$

$$
\int_{b<r}d^3xe^{i\vec q\cdot\vec x}\frac{x_{\alpha}x_{\beta}}{r^4}=\frac{4\pi}{q}\{\frac{q_{\alpha}q_{\beta}}{q^2}[\frac{-3\sin}{2a^2} 
+\frac{3\cos a}{2a}-\frac12\int_a^{\infty}\frac{\sin u}{u}du]+
$$
$$
\delta_{\alpha\beta}[\frac{\sin a}{2a^2}-\frac{\cos a}{2a}+\frac12\int_a^{\infty}\frac{\sin u}{u}du ] \},
                                                                                            \eqno(46)
$$
$$
\int_{b<r}d^3xe^{i\vec q\cdot\vec x}\frac{x_{\alpha}x_{\beta}}{r^5}=4\pi\{\frac{q_{\alpha}q_{\beta}}{q^2}[-\frac{\sin a}{a^3} 
+\frac{\cos a}{a^2}]+
$$
$$
\delta_{\alpha\beta}[(\frac{1}{3a}+\frac{1}{3a^3})\sin a-\frac{\cos a}{3a^2}+\frac13\int_a^{\infty}\frac{\sin u}{u}du ] \},
                                                                                            \eqno(47)
$$
Putting $\alpha=\beta$ in (46) and (47) we get
$$
\int_{b<r}d^3xe^{i\vec q\cdot\vec x}\frac{1}{r^2}=\frac{4\pi}{q}\int_a^{\infty}\frac{\sin u}{u}du, \eqno(48) 
$$
$$
\int_{b<r}d^3xe^{i\vec q\cdot\vec x}\frac{1}{r^3}=4\pi\left(\frac{\sin a}{a}+\int_a^{\infty}\frac{\sin u}{u}du\right).
                                                                                              \eqno(49)
$$

Now it is easy to verify that the sum of contributions from $r<b$ and $r>b$ to Fourier-transforms of differences in (39) and (40) do not contain terms proportional to $\delta_{\alpha\beta}$. Similar statement is true for $h_{\alpha\beta}^{(2)har}-h_{\alpha\beta}^{(2)iso}$.
This means that if two $h_{\alpha\beta}^{(2)} $ differ only by gauge function then they lead to the same scattering.

So, it is shown that $h_{\alpha\beta}^{(2)priv}$ with term (28) leads to the same scattering  as
$h_{\alpha\beta}^{(2)har}$.
\section*{References}
1. V.A.Fock, The theory of space-time and gravitation, Pergamon, New York,1964. \\
2.  S.Weinberg, Gravitation and Cosmology,John Wesley, New York,1972.  \\
3. R.Feynman, F.Moringo, W.Wagner, Feynman Lectures on Gravitation. Addison-Wesley.     \\
4. A.I.Nikishov, Gravity from the viewpoint of theory of sources, arXiv:1605.06305v1 [physics.gen-ph] 16
May 2016  \\ 
5. N.V. Mitskevich, JETP, {\bf 34}, 1656 (1958) (In Russian).\\
6. A.I.Nikishov, Gravity according to theory of sources, arXiv:1612.04198 v.1, (physics gen-ph) 7 Dec 2016. \\
7. A.I.Nikishov, On the simplified tree graphs in gravity.  arXiv:1111.0812 (physics gen-ph).(2011).\\  8. L.Landau and E.Lifshitz, The classical theory of fields, Oxford,Pergamon Press,1983.                                                                           
\end{document}